\documentstyle[12pt]{article}
\date{}

\newcommand{\beq}{\begin{equation}}
\newcommand{\eeq}{\end{equation}}
\newcommand{\beqn}{\begin{eqnarray}}
\newcommand{\eeqn}{\end{eqnarray}}
\newcommand{\gtrsim}{\stackrel{>}{\sim}}
\newcommand{\lessim}{\stackrel{<}{\sim}}

\newcommand{\dst}{\displaystyle}

\newcommand{\fr}[2]{\frac{{\dst #1}}{{\dst #2}}}

\newcommand{\intlum}[1]{{\rm annual luminosity} $ #1$  fb$^{-1}\,$}

\newcommand{\mww}{\mbox{$M_W^2\,$}}
\newcommand{\mh}{\mbox{$M_H\,$}}
\newcommand{\mhh}{\mbox{$M_H^2\,$}}
\newcommand{\mz}{\mbox{$M_Z\,$}}

\newcommand{\sigmaw}{\mbox{$\sigma_W\,$}}

\newcommand{\sww}{\mbox{$\sin^2\Theta_W\,$}}

\newcommand{\ptr}{\mbox{$p_{\bot}\,$}}
\newcommand{\ptrs}{\mbox{$p_{\bot}^2\,$}}
\newcommand{\lgam}{\mbox{$\lambda_{\gamma}$}}

\newcommand{\lel}{\mbox{$\lambda_e$}}
\newcommand{\epe}{\mbox{$e^+e^-\,$}}
\newcommand{\ggam}{\mbox{$\gamma\gamma\,$}}
\newcommand{\egam}{\mbox{$e\gamma\,$}}
\newcommand{\gewnu}{\mbox{$e\gamma\to W\nu\,$}}
\newcommand{\eeww}{\mbox{$e^+e^-\to W^+W^-\,$}}
\newcommand{\ggww}{\mbox{$\gamma\gamma\to W^+W^-\,$}}
\newcommand{\ggzz}{\mbox{$\gamma\gamma\to ZZ\,$}}
\newcommand{\geeww}{\mbox{$e\gamma\to e W^+W^-\,$}}

\newcommand{\pair}[1]{\mbox{$#1 \bar{#1}$}}

\title{PHYSICAL PROGRAM FOR PHOTON COLLIDERS\\
{\normalsize\em Talk given at 10 Workshop on Photon--Photon
Collisions, Sheffield, April 1995}}

\author{Ilya F.~Ginzburg\\ Institute of Mathematics.
  630090. Novosibirsk. Russia.\\ E-mail: ginzburg@math.nsk.su}

\begin{document}

\maketitle

\begin{abstract}
{The physical program for the photon linear colliders (PLC) is
discussed.  We consider the problems in electroweak theory,
hadron physics and QCD, new particles and interactions. We
arrange this program for the main stages of PLC, which are the
parts of an entire program for linear colliders. Some important
unsolved problems are listed.}
\end{abstract}

\section{ Main expected features of photon colliders}

We believe now that

\begin{quotation}

{\bf Most of future linear colliders (LC) will be \epe/\egam/\ggam complexes.

The entire project of a LC should have an ambitious goal, e.g.,
to achieve energy about 2 TeV. Its total realization will be a
 long and expensive task. However the specific features of LC's
provide an opportunity to do this job step by step, depending on
present day reserves of funds and time. The stops along the way
will be the separate colliders with ever higher beam energies and
with new and higher physical potentials.}
\end{quotation}
\vspace{0.1cm}

The idea of Photon Linear Colliders (PLC) based on \epe\ LC was
proposed and developed in papers \cite{GKST}.  Some details of
design and some separate phenomena in PLC were considered in the
subsequent papers [2--5]. The relation to the modern projects of
LC's is discussed in the Telnov's talk at this Workshop
\cite{Tel95}.

We will denote below: $E$ -- initial electron energy, $E_{\gamma}
$ -- photon energy, $E_{\ggam} $ or $E\egam$-- c.m. energy of
photon--photon or electron--photon system.

Modern studies show that one can expect the following parameters
of PLC:

\begin{itemize}
\item Characteristic photon energy is $E_{\gamma } \approx
0.8E$.
\item Mean energy spread is $<\Delta E_{\ggam}/ E_{\ggam}>
\approx 0.1$  ({\em monochromatic variant}).
\item Photons will be polarized, with mean helicity
$<\lambda>\approx 0.95 $.
\item There is no special physical reasons for observations at
small angles, except details of design.
\item One can expect \intlum{10\div 20} in the {\em monochromatic
variant}. It is about $20\%$ from that for the basic \epe LC.

This value can be obtained on the basis of modern LC projects,
optimized for the \epe mode. If one optimize these projects from
beginning for \ggam mode, one can hope for luminosities, which
are higher by an factor\footnote{ I am grateful to V.Balakin and
A.Skrinsky who clarify me this opportunity.} $\gtrsim 10$.
\item In each case one can organize {\em nonmonochromatic variant}
with \ggam luminosity $\sim 5$ times higher with wide energy
spectrum (and with almost the same high energy part of spectrum
as in the monochromatic variant). The additional, more soft
photons are almost unpolarized in this case.
\item One can organize, in principle, the {\em supermonochromatic
variant} with $E_{\gamma }\approx 0.95E;\quad <\lambda>\approx
0.95\;$; $ <\Delta E_{\ggam}/ E_{\ggam}> \approx 0.015\div 0.02$
but with luminosity which is about $10\div 20$ times
less\footnote{ Variation from monochromatic variant to the
nonmonochromatic one seems to be possible during operations;
going to the supermonochromatic variant demands for new optical
technique.}.
\end{itemize}

\begin{quotation}
{Besides,{\em the conversion region is (very nonsymmetrical) \egam
and \ggam collider with a small c.m. energy ($E_{\egam} \approx
1.2 $ MeV, $E_{\ggam} \approx 1$ MeV), but with a huge luminosity
$10^6\div 10^8 \mbox{ fb}^{-1}$ per year.} It provides new
opportunities to hunt for the very light particles \cite
{Pol,GKP}.}
\end{quotation}

We discuss below physical program for PLC.  The separate programs
for different stages of PLC (PLC0 -- PLC3) are considered briefly
in the end of the text. Some stages of such program for \epe LC's
were considered in refs. \cite{KEK1}--\cite{Haw}. Some
points related to PLC have been discussed in refs.
\cite{Sar}--\cite{SD},\cite{LBL}.

\section{ Hadron Physics and QCD}

The hadron physics and QCD are the traditional problems for two
photon experiments. These experiments provide new type of
collisions and with the simplest quark structure of the pointlike
initial state. The PLC's will spread these studies to quite new
regions. Here these studies will be no less informative than
those at other colliders. The results from PLC's together with
those from the Tevatron and HERA, will produce the entire set of
complementing data related to a factorized (in the old Regge
sense) set of processes. In this respect, HERA acquires a new
importance of a bridge between PLC's and Tevatron/LHC.\\

{\large\bf Basic problems in Hadron Physics and QCD at
PLC}( In more details see, e.g., \cite{GinSD,GinLBL}.).\\

1) TOTAL CROSS SECTION {\large $\sigma_ {\gamma\gamma\to
hadrons}$}. It is important to study the energy dependence of
this cross section (together with the $Q^2$ dependence --- in
$\gamma e$ collisions). Its comparison with $\sigma_{pp}
(\sigma_{p \bar p})$ and $\sigma_{\gamma p}$ will allow us to
understand the nature of the increase in the hadron cross
sections with energy. The crucial problem is to test the possible
factorization of these cross sections (this factorization is
assumed in ref.\cite{CBP}). (This cross section is expected to be
$\sim 0.3\;\mu b$ in the SLC energy region, and $\sim 0.5\div
1\;\mu b$ at $E_{\ggam} \sim 2$ TeV \cite{CBP}).  How can we
measure this cross section?

2) The same questions can be posed for the DIFFRACTION LIKE
PROCESSES IN SOFT REGION with production of neutral vector mesons
V (Pomeron) and pseudoscalar P or tensor T mesons (Odderon). That
are the processes
\beq
\ggam\to M+M'; \; \ggam \to M+X \mbox { with rapidity gap}
\quad (M=V \mbox { or } T/P).
\label{dif}
\eeq
(X is the hadronic system separated from meson M by large
enough rapidity gap).

The cross section of process $\ggam\to \rho^0\rho ^0$ is
expected to be $\sim 0.1\sigma _{tot}$ (see \cite{BGMS}).  The
processes $\ggam \to\pi^{0}\pi^{0},\;\ggam \to\pi^0a_2,\ldots $
are described only by Odderon exchange. These present a unique
opportunity for Odderon study. The real role of Odderon in
hadron collisions is unknown today. This is in contrast with
the fact that the Pomeron and the Odderon have the identical
status in pQCD. But where is the Odderon in the data?

3) PHOTON STRUCTURE FUNCTION is discussed widely at this
Workshop. At PLC it will be studied in new region and with high
accuracy (target energy is roughly known). Modern studies show
that the real influence of the hadronic component of the photon
is very important over a wide region of parameters \cite{sfWit}.
It is expected that at the higher values of $Q^2$ the pointlike
component of photon \cite{Wit} becomes dominant. The small $x$
region is related to the Pomeron in pQCD (perturbative Pomeron)
\cite{BFKL}.

\begin{table}[hbt]
\begin{center}
\caption{\em Positions of different colliders in discussed problems
of hadron physics}
\vspace{0.3cm}

\begin{tabular}{|p{6cm}||p{2cm} |p{2cm} |p{2cm} |p{2cm}||}\hline
&PLC&HERA, LEP+LHC& Tevatron, LHC&\epe LC\\ \hline
nature of $\sigma_{tot}^{had}$ growth &\multicolumn{3}{|c|}{
Comparison}& --- \\ \hline
Pert. Pomeron and odderon&\multicolumn{2}{|c|}{ the best}
& ? & --- \\ \hline
minijets nature&\multicolumn{3}{|c|}{Comparison}& --- \\ \hline
Photon structure function& the best& --- & --- &second\\ \hline
large angle jets for discovery of New Physics& second
&\multicolumn{2}{|c|} { specific}&the best\\ \hline
t quark, threshold phenomena& second& -- & -- & the best\\ \hline
t quark, properties& the best& --- & --- & second\\ \hline
\end{tabular}
\end{center}
\end{table}

4) Three mechanisms of LARGE ANGLE JET production
\cite{GISHaw} are substantial:

QUARK EXCHANGE is described by the simplest pQCD diagram with
two quarks (jets) in the final state. The corresponding cross
section is $d\sigma^q\propto 1/(s\ptrs)$.

Diagrams with GLUON EXCHANGE between two pairs of quark (jets)
give the cross section $d\sigma^g\propto 1/(\ptrs)^2$.

The W PRODUCTION mechanism is \ggww reaction with subsequent
decay of W's into quark jets (at $\sqrt {s}>160$ GeV).

At $\ptrs\sim s$ W production dominates, $d\sigma^W> d\sigma^q
>d\sigma^g$. When \ptr decreases, the relative value of quark
exchange decreases as $p_{\bot}^2/s$ in comparison with other
mechanisms and gluon exchange become dominant. (In the \egam
collisions W production through process \gewnu provides dominant part
of produced quark jets for almost all \ptr.)

The NEW PHYSICS can manifest itself as anomalous large angle jet
production, obliged by anomalous interaction like $\ggam q\bar
q$.

5) THE SEMIHARD PROCESSES are those, for which the characteristic
value of transverse momentum is small in comparison with total
energy but large in comparison with the strong interaction scale
$\mu\sim 300\mbox{ MeV}:\quad s\gg \ptrs\gg\mu^2$ (THE SMALL
ANGLE JET PRODUCTION, THE DIFFRACTION LIKE PROCESSES (\ref{dif}),
THE PHOTON STRUCTURE FUNCTION AT SMALL x, etc.). These phenomena
provide us information about perturbative Pomeron and odderon,
mechanisms of shadowing in pQCD, etc.

In this region, a new parameter appears in the pQCD series,
$\alpha_s(p_{\bot}^2) ln(s/p_{\bot}^2)\approx\alpha_s(p_{\bot}^2)
\eta$ or $\alpha_s(Q^2) ln(1/x)$, that becomes large while s
increases. Therefore, the entire pQCD series should be taken into
account, and studies here provide opportunity to test the inner
structure of pQCD in all orders. Due to the simple pointlike
nature of photons, the nontrivial results in pQCD could be
obtained almost without model assumptions. Unfortunately, the
influence of hadronlike component of photon is expected to be
relatively small at large enough \ptr only. For example, for the
diffraction like processes (\ref{dif}) it is expected to be at
$\ptr> 7$ GeV \cite{GIS95}\footnote{
\normalsize The cross sections of some processes, integrated over
this range of \ptr and with large enough rapidity gap, are
estimated from below as: $\sigma_{\ggam \to \rho^0 X}\gtrsim 1$
pb, $\sigma_{\ggam \to\pi^0 X} \gtrsim 0.4$ pb \cite{GPS,GIv}.
The first quantity should be multiplied by the growing BFKL
factor (see \cite{Muel,Iv}).}.

Where is the corresponding boundary for the jet production with
rapidity gap?

Where are the real bounds for the description of $s/\ptrs$ dependence
with perturbative Pomeron or odderon?

What is the nature of shadowing effects?

6) MINIJETS. The number of minijets within some definite
angular interval (e.g., $1^{\circ} - 179^{\circ} ,\;5^{\circ} -
175^{\circ} $) should be weakly dependent on energy
\cite{GISHaw}. Hence, the energy dependence of transverse energy
flow will show us the energy dependence of mean transverse
momentum for single minijet.

7) The detailed study of {\large t} QUARKS is a problem for LC's.

The threshold effects will be investigated both at \epe500 and
its \ggam modification \cite{FKhoz}. The strong dependence on
photon helicities provides opportunity to see delicate details of
\pair{t} interaction near the threshold.

The PLC provides the best opportunity for study of t--quark
properties themselves. Indeed, the cross section of $t\bar t$
production in the \ggam collision is larger and it decreases more
slowly with energy than that in \epe collision (Fig. 2).
Therefore, relatively far from the threshold one can expect at
PLC about $10^5$ $t\bar t$ pairs/year, and their decay products
are overlapped weakly. Some rare $t$--decays could be studied
here.

\section{ Higgs Boson Physics}

1) THE DISCOVERY OF HIGGS BOSON (Higgs) seems to be the most
important problem of modern particle physics. The result of
competition between different colliders depends on both Higgs
mass and the time of beginning of operations there. In any case,
the SM Higgs with $M_H\leq 90$ GeV will be either found or
prohibited at the LEP2.

If $M_H > 2M_Z$, the Higgs will be detected at the LHC in the
decay mode $H\to ZZ $. This Higgs will be seen at PLC2 through this
very decay mode if its mass is less than $400\div 500$ GeV. (The
upper bound is obliged by 1) compensation of t quark and W boson
loop contributions into the Higgs two photon width and 2) large
background due to \ggzz process \cite{Jikzz}).

The PLC1 seems to be the best machine for the discovery of Higgs
with mass $80\div 180$ GeV \cite{BalG}\footnote { The modern
activity about observation of such Higgs at the LHC shows very
questionable possibilities here. The $e^+e^-300$ LC will covers
mass interval below 150 GeV \cite{Zer}, but it will be built
later than PLC1.}.  The process $\ggam\to H\to b\bar b$ with QED
background $\ggam\to b\bar b$ was considered in \cite{BBC}, these
results are refined in ref. \cite{Khhiggs,JikT}. It is follows
 from these results, that with using of monochromatic variant of
PLC with zero total initial photon pair helicity, one can observe
Higgs with mass $80 <M_H<150 $ GeV, based on the luminosity
integral about 3 fb$^{-1}$.

The mass interval 150 GeV $<\mh<2\mz$ is the most difficult one for
the Higgs discovery. Indeed, the decay $H\to WW$ dominates here
strongly, but the WW production cross section via Higgs is much
less than that without this intermediate state. The new
opportunity follows from the recent result \cite{MTZ}.
It was noted there, that the total width of the heavy Higgs
will by high enough to resolve details of WW spectrum within
this width interval. Besides, the amplitude of the $\ggam \to
H \to WW$ process is complex with the phase, which varies fast:
$$
M\propto \Gamma(\ggam\to H)\fr{1}{s-\mhh+i\Gamma_H\mh}
\Gamma(H\to WW)
$$
(The first factor here is also complex.) Therefore, the interference
of this amplitude with that for QED process \ggww\  is high enough.
The paper \cite{MTZ} shows the spectacular curves for 180 GeV
$< \mh <400$ GeV\footnote{ The amplitude of this interference is
higher at lower $s$, since W's from Higgs decay are polarized
longitudinally and the fraction of longitudinal W's from \ggww
process decreases with s.}. The special simulation work is
necessary to understand, how are requirements imposed on either
PLC (monochromatization degree) or detector (accuracy of W decay
products momenta measurements), to see Higgs in the widest mass
interval.

2) At the PLC only, one can measure precisely the HIGGS TWO
PHOTON WIDTH. This width is the counter for SM particles heavier
than a Higgs.

\begin{table}[htb]
\begin{center}
\caption{\em Higgs physics at various colliders}
\vspace{0.3cm}

\begin{tabular}{|p{3cm}|p{3cm}|p{2.5cm}|p{2.5cm}|p{2cm}|}\hline
&PLC1, $\sqrt{s}<180$ GeV&PLC, next stages&\epe LC& LHC\\ \hline
$80<M_H<150$& discovery ($b\bar b$), ?--study ($\tau \bar{\tau}$)&
-- & After PLC1 discovery& ??\\ \hline
$150<M_H<180$&discovery ($WW$)?& -- & ?? & --\\ \hline
$180<M_H<400$& -- & study & & discovery\\ \hline
$400<M_H<800$& -- & If anomalies?& & discovery\\ \hline
$H^3$ vertex& -- & weakly& good& ??\\ \hline
anomalies in Higgs sector& $\pm$& good& + & accuracy?\\ \hline
\end{tabular}
\end{center}
\end{table}

3) The investigation of HIGGS COUPLING WITH MATTER is necessary to
obtain whether the observed particle is actually a Higgs of
the SM or something else. It will be an essential test for the
Higgs nature of quark masses and (as some alternative) possible
SUSY Higgs.

If $M_H< 150$ GeV, one could investigate Higgs decay into
\pair{\tau} or \pair{c} with SM branching ratios $\sim 0.06 \mbox
{ or } 0.04$ (cf.\cite{JikT}). These opportunities need for new work with
simulation.

If $\mh> 2\mz$, one can compare Higgs coupling with Z and W (by
comparison of Higgs production via reaction $\ggam\to H\to ZZ$
and via interference in \ggww reaction).

If $M_H\sim 2M_t$, the interference between QED process $\gamma
\gamma \to t\bar t$ and resonant one  $\gamma \gamma \to H\to
t\bar t$ can be used to see the value of Higgs coupling with
t-quark \cite{BGMel}.

3) THE ANOMALOUS INTERACTIONS OF HIGGS The SM Higgs with $M>500$
GeV will be invisible in a photon--photon collision. Therefore,
any Higgs signal at a PLC in this region manifests the existence
of either some heavier SM particles or nonstandard interactions
of Higgs, having the scale about a few TeV \cite{GHiggs,GinSD}.

\section{ Gauge boson physics}

Modern data show that the Standard Model is the theory of our
world. Meanwhile both the checkings up of this theory in new areas
and the observation of some (now unknown) features seem to be
necessary. After LEP study of Z peak, photon colliders will be
the best laboratory for this aim.

The sketch of THE MAIN PROCESSES WITH W AND Z PRODUCTION at PLC
within SM is given in refs. \cite{Wrev,BBG}. The scale of these
phenomena at PLC is given by the quantity $\sigmaw =8\pi\alpha^2
/\mww \approx 81$ pb, which is the cross section of \ggww\
process at high enough energies. This very process (together with
\gewnu) determines PLC as {\large\em W factory} with $10^6\div 10^7$
W's per year.

The cross sections of processes \gewnu and \ggww with their dependence
on helicities of photon \lgam and electron \lel are considered in ref.
\cite{GKPS}. The \egam cross sections at $E\egam <200$ GeV and
the \ggam\ cross sections at $E\egam <300$ GeV are varied
strongly with variation of photon helicities. The more detail
study of this dependence taking into account finite W width is
desirable. Perhaps, this provides additional opportunity to see
the admixture of gauge boson interactions or decays which violate
CP invariance. This dependence on photon polarization vanishes
at large enough energies.

Besides, the process \gewnu is switched on or off entirely with
variation of electron helicity ($\sigma_{\gewnu}$ $= (1-2\lel)
(\sigma +\lgam \tau)$). This means, that this process is
very sensitive to admixture of right--handed currents in $W$
coupling with matter.

The angular distribution of produced W's for both processes is
more favorable for W recording than that in process \eeww. The
\ggww cross section tends fast to its asymptotic value \sigmaw.
At large enough energies we have $\sigma_{\gewnu} = \sigmaw /8\sww
\approx 43$ pb.

When the energy increases, the cross sections of a number of
higher--order processes become large enough. The catalogue of
such processes of third order in SM was obtained with CompHEP
package \cite {CompOur}, see Fig. 1. I will don't
enumerate all papers, devoted to separate processes here.

The different processes (and different kinematical regions for
one process) are sensitive in different manner to various
possible anomalous gauge boson interactions. Therefore, the
comparative detail study of different processes provides
opportunity to study various anomalies separately. The special
work is necessary to present detail program in this field.

Among the processes of highest interest is the process \geeww
with high cross section. The large enough fraction of this cross
section is given by region, which is very sensitive to the
$\gamma ZWW$ interaction \cite{geeww}.

In fourth order processes we can see subprocesses with heavy
gauge boson scattering. The SM cross sections for the processes
$\ggam\to WWWW$ and $\ggam\to WWZZ$ are $\sim 0.3\div 0.1$ pb
\cite{Jik4}. The cross section of the process $e\gamma \to eWWZ$
is of the same order of value:
$$
(\alpha/\pi)^2 \ln(s/m_e^2)\cdot\ln^2(s/4\mww)\sigma_{\ggww}.
$$

Some process of fifth and sixth order will be observable at high
enough energies, for example, $\egam to\epe e WW$, $\egam \to
\epe \nu WZ$, $\ggam\to \epe \pair{\mu} WW$, etc.

\begin{table}[htb]
\begin{center}
\caption{\em Gauge boson physics at different colliders}
\vspace{0.3cm}

\begin{tabular}{|p{8cm}|p{7cm}|}\hline
&PLC will be W factories with $\sim 10^6\div 10^7$ W's per year\\
\hline
parameters of W& LHC (only $M_W$)$\to$ LEP200 $\to$ PLC\\ \hline
right hand currents in $W\to e\bar{\nu}$& PLC1 $\to$ PLC
(\gewnu)\\ \hline
multiple W production &PLC better than \epe LC and LHC \\ \hline
Testing of deep properties of QFT (quantization of theory with
unstable particles)&PLC only (necessary accuracy
$\sim\alpha(\Gamma/M) \sim 10^{-3}$)\\ \hline
CKM matrix elements measuring& PLC better than \epe LC\\ \hline
QCD effects in WW interactions & PLC only\\ \hline
Testing of EW SM perturbation theory, $\alpha\ln^2(s/\mww)$
summation& PLC2 TeV only\\ \hline
\end{tabular}
\end{center}
\end{table}

\subsection{Problems in gauge boson physics}

{\large\bf\em Insertion of finite W width}\\

To describe gauge boson production with real final states of the
$W$ decay, one should use the $W$ propagator near its physical
pole. To avoid divergence, it is necessary to insert in this
propagator the $W$ width $\Gamma_W$, for example,
\beq
{1\over k^2-M_W^2-i\varepsilon} \Rightarrow {1\over
k^2-M_W^2-iM\Gamma_W}. \label{Wprop}
\eeq

This simple substitution can violate gauge invariance \cite
{ACOld} and unitarity. The naive recipes used here results in
inaccuracy's $\sim (1\div 3)\Gamma/M$. The more likely recipes
(which unknown until now) should eliminate the above violations.
Nevertheless, this requirement gives no unambiguous recipe. One
can expect that the ambiguity of the result when using the
different recipes, both unitary and gauge invariant, without
genuine theory will be $\sim\alpha \Gamma/M$ ($\sim 10^{-3}$ or
larger), i.e. the accuracy of such recipes seems to be deficient
for description of data. Therefore, the well--known fundamental
problem of quantum field theory becomes of practical importance
here\footnote{ It appears here first as a practical one in
particle physics. Indeed, the same problem for $Z$ peak is solved
for the unique object produced without other $W$'s or $Z$'s;
therefore in this case the gauge invariant recipe is sufficient.
The description of last stages of hadron reactions contains some
phenomenological models; therefore, any uncontradictory recipe is
suitable here. For $\mu$ and $\tau$, the $\Gamma/M$ ratio is less
than the corresponding coupling constant $\alpha$.} (see e.g.
\cite{Velt}):

\begin{quote}
{\bf It is necessary to construct a genuine theory of unstable
gauge bosons}. The key place is in the quantization of unstable
fields.
\end{quote}
\vspace{0.3cm}

{\large\em Radiative corrections (RC)}\\

To understand the future data from PLC's, the RC to the discussed
cross sections should be calculated. These RC can be subdivided for
3 types: those caused by finite width of W and Z, those caused
by initial state radiation, and the "proper" RC.

{\em (i)} Corrections caused by the finite width of W or Z.
Finite width results in some enlargement of the final phase
space. This enlargement can be used to extract some parameters of
W more precisely. For example, one can use the $e^+e^- \to WW$
process below threshold to obtain more precise values of the
width and mass of W \cite{GKPS,Wthr}. The above discussion shows
the ambiguity in such description of these parameters.

{\em (ii)} The RC, caused by the initial state radiation in the
\egam collisions are often calculated by the method from ref.
\cite{4}. These very large nonspecific RC can be treated as the
nonmonochromaticity of the initial electron energy \cite {Coll}.

{\em (iii)} The "proper" RC due to electroweak interaction are of
more interest for they are sensitive to additional heavy
particles or violations of the SM (cf. \cite{AP}).

In the study of real processes it is necessary to calculate the
RC due to interaction of W decay products. The largest RC of this
type are related to the strong interaction between quarks (jets)
 from the decay of different W's. The Pomeronlike effects can
intensify these RC at $s\gg M_W^2$.\\

{\large\bf\em Some other problems at PLC's}\\

Let us list some problems in gauge boson physics, which was not
discussed above.

1) The underlying interactions could manifest itself as the
deviations from the SM in some ANOMALOUS INTERACTIONS OF GAUGE
BOSONS. These anomalies are described by some effective
Lagrangian. The standard approach is to consider here operators
of lower dimension -- 4 and 6 (e.g. an anomalous magnetic moment,
quadruple moment, etc.). Usually, these effects increase with
energy, the larger energy is the better for their detection. Some
results had been obtained for the $e^+e^-500$ LC (including PLC)
\cite{26,27,43a}.

The values of anomalous coupling constants, which can be
considered, are limited from below by condition, that the
anomalous contribution into cross section should be larger than
RC in SM (see Appendix for more details). At PLC we expect to
meet these bounds.

2) One can look for THE POSSIBLE $CP$ VIOLATION IN THE $W$ BOSON
DECAYS, e.g., $ \hbox {BR}(W^-\to e^-\bar \nu)\neq \hbox
{BR}(W^+\to e^+\nu$ )\footnote {The value of this effect in the
"most optimistic" case can run into $ 10^{-3}$ \cite{CP}.} In
this problem, the additional opportunities should be considered,
which given by the initial photon state with total helicity 0.

3) One can measure THE ELEMENTS OF THE
CABIBBO--KOBAYASHI--MASKAWA MIXING MATRIX on the mass shell of
$W$. Their comparison with those obtained in the update
experiments (far from $W$ mass shell) can give an idea about
their dependence on $W$ boson virtuality.

4) The STRONG INTERACTION IN HIGGS SECTOR seems to be very
probable one at $\sqrt{s}\gtrsim 1$ TeV independent on Higgs mass,
since the interaction Higgs t quark is strong one. It could
manifest itself at PLC3 as new effects in gauge boson
interactions, like resonances in the gauge boson systems, unusual
energy dependence, multiple W production, etc. Its first signals
could be obtained in the production of longitudinal W's and Z's.
(For more details, see a number of papers, e.g., \cite{33a}).
Therefore, this is a problem for gauge boson physics too.

The SM cross section $\gamma \gamma \to Z_L Z_L$ is small
\cite{Jikzz}. However, experience in pion physics permits us to
expect here the large effects due to some heavy states (like
$\omega$ in the $t$-channel for $\gamma \gamma \to
\pi\pi$). The relatively large contribution from the transverse
W loop has no analogy in the pion world.

At the large enough energies, one can expect to see the strong
interaction of transverse W's driven by the strong Higgs
self--interaction. Where does this energy region begin?

\section{New Physics}

Two opportunities are considered, when we speak about New Physics
effects --- the discovery of NEW PARTICLES and NEW NONSTANDARD
INTERACTIONS of known particles.

PLC's provide the best place to discover many new particles ---
in comparison with other colliders, having similar energy. The
reasons for this statement are:

1) The signal to background (S/B) ratio at PLC is often much
better than that at hadron colliders (see in more detail ref.
\cite{GinLBL}).

2) The photons are "democratic" respective to all charged
particles. Therefore, the analyses of new particles production
have no additional ambiguities due to production mechanism at PLC
(which exists in collisions with hadrons)\footnote{ One should
remind in this respect that a number of modern limitations to
masses of new particles were obtained with some additional
hypotheses, which are often not so natural. For example, the
modern limitation to a mass of leptoquark from HERA data was
obtained, assuming its coupling with electron and
quark,, is not small. There are models, in which
this $G_{\ell q}$ is of order of ratio of sum of mass,
consisting leptoquark to a Higgs v.e.v. $v\approx 250$ GeV:
$G_{\ell q}\sim S_{\ell q}(m_{\ell} + m_q) /v$. Here the $S_{\ell
q}$ is the corresponding mixing matrix element.  Therefore, the
coupling of electron with some u or d or s quark should be very
weak in such a model, and the corresponding leptoquark could not
be seen at HERA even if it is light enough.)}.

3) The cross sections of charged particles production at \ggam
PLC are larger than those at \epe LC (see Fig. 2). Even if PLC
luminosity is 5 times less than that for basic \epe LC (standard
monochromatic variant), the number of produced pairs at \epe
collider is no greater than that at \ggam collider. Besides, this
production in \ggam collision decrease with energy more slow than
that in \epe collision. This provides an opportunity to study new
particles relatively far from threshold with a good enough rate.
In this region, the decay products of these new particles are
overlapped weakly.  Therefore, theirs detailed study should become
more feasible.

4) The \ggam collisions produce often the pairs of identical
particles with identical decays (e.g., $\ggam\to\pair{\tilde{\mu}
}$). This makes easier the analysis of events with missed \ptr.

5) In contrast with hadrons, a photon is pointlike, its
quark content is well known. Entire photon energy is used to see
the small distance phenomena of interest.

6) In some cases \egam collisions are preferable (for example,
reactions $\egam\to e^*$, $\egam \to W\nu_e^*$, $\egam \to \tilde
e \tilde \gamma$.)

On the contrary, gauge invariance is constrained strongly
interactions of matter with photons. Therefore, the effects of some
new interactions are suppressed here. On the other hand, it means
that the origination of observed effects would be separated
easily.

The discovery potential of PLC for new particles and interactions
was considered in numerous papers (see also reviews \cite{GinSD,26}):

\begin{itemize}
\item SUSY particles \cite{Cuyp}:
$$
\tilde{W^{\pm}},\;\tilde{Z}:\;\;\egam\to \tilde{W}\tilde{\nu};\;
\ggam\to \tilde {W^+}\tilde {W^-};\; \egam \tilde{Z}\tilde{e};\;\;
\tilde{P}\equiv\tilde{\ell},\;\tilde{H^{\pm}},\;\tilde{u},\;\tilde{d}:\;
\ggam \to \tilde{P}\tilde{P}.
$$
\item Excited leptons and quarks \cite{GIvBoud}:
$$
e^*:\;\; \egam\to e^*;\quad \ell^*:\;\;
\ggam\to\ell^*\bar{\ell}; \quad
\nu^*:\;\;\egam\to \nu^* W;\quad q^*:\;\;\ggam \to
q^*\bar{q}.
$$
\item Leptoquarks \cite{lq}.
\item Charged Higgses.
\item Composite scalars and tensors.
\item Dirac--Schwinger monopoles with mass $\lessim 10 E$ \cite{GPmon}.
\item Invisible axion (from the conversion region) \cite{Pol,GKP}.
\item Higgs nonstandard interactions \cite{GHiggs}.
\end{itemize}

\section{Main stages of PLC program}

To see for discussed problems from more practical point of view,
we consider briefly the opportunities of various stages of entire
program for LC. We will point in this list the points only with
the known now energy scale. Certainly, one should study hadron
physics and hunt for new particles and interactions at all PLC's.

\subsection{ PLC0 --- Photon collider, based on SLC.}

Stanford Linear Collider (SLC) -- the sole Linear Collider, which
works now, can provide a realistic test bed for a high energy PLC
\cite{MemLBL}. The characteristic energy for such -- preliminary
stage -- PLC will be $E_{\ggam}\approx 70\div 80$ GeV, and its
annual luminosity can be $\sim 1$ pb$^{-1}$ (or $\lessim 100$
pb$^{-1}$ if rather large variations in SLC design will be
realized\footnote{ I am grateful to V.~Balakin, who tell me about
this opportunity.}).

These numbers show that such collider could be only used for the
study of hadron physics and QCD. It will be the substantial step
in comparison with modern studies at \epe colliders (which were
discussed at this Workshop). The specific problems for this PLC
are (\cite{GinSD}):

TOTAL CROSS SECTIONS, DIFFRACTIVE PROCESSES, PHOTON STRUCTURE
FUNCTION, JETS.

To use the potential of this PLC as whole as possible, the
small angle detector is necessary.

\subsection {PLC1 --- with $e_{\ggam}\leq 180\div 190$ GeV}

This stage PLC was proposed in \cite{BalG}. It needs no positron
beams, since this energy interval will be covered by LEP2.

1) HIGGS HUNTING AND STUDY will be the central idea of this stage
PLC.

The discovery of Higgs boson with $M_H<150$ GeV via $\ggam\to H
\to b \bar b$ reaction at a PLC in a monochromatic variant
demands about 3 fb$^{-1}$ integrated luminosity
\cite{BBC,Khhiggs}. Perhaps, it is possible to see Higgs decay to
\pair{\tau} or \pair{c} in supermonochromatic variant? Perhaps,
it is possible to discover Higgs with mass $150<\mh<180$ GeV via
interference of $\ggam\to H\to WW$ process with QED process
\ggww \cite{MTZ}?

2) In the process \gewnu (with $10^4\div 10^5$ W's per year --
depending on beam energy) we will see the SINGLE W.  The
(Coulomb) interaction with other produced particles will be
absent. In the almost all channels the interference with nonpole
diagrams will be very small, i.e., problems with gauge invariance
at incorporation of total width will be also absent.  Therefore,
these experiments will give us more precise and unambiguous values
for W PARAMETERS.

This process is switched on or off with variation of basic
electron helicity. Therefore, it provides the best place for THE
TESTING OF ADMIXTURE OF RIGHT--HANDED CURRENTS IN THE $We\nu$
VERTEX (in quite other point than modern $\mu$ decay).

3) The \ggww process near the threshold depends strongly on
photon helicities. Its study with finite W width should be made.
Perhaps, these polarization phenomena are sensitive to some
anomalous gauge boson interactions?

\subsection{ PLC2 --- based on \epe500}

1) This stage PLC begin new era --- era of PLC as a SPECIAL
COLLIDER FOR A STUDY OF GAUGE BOSON PHYSICS.  Both \ggam and
\egam PLC's become W factories with about $10^6\div 10^7 $ W's
per year. The entire set of problems related to EW SM bosons will
be solved here with high precision.

2) The study of {\large t} QUARK PRODUCTION near the threshold (in the
states with different polarizations).

3) HIGGS HUNTING AND STUDY if $M_H<400$ GeV.

\subsection{ PLC3 --- with energy up to 2 TeV}

1) The main problem for SM physics here are the study of both the
basic processes \ggww , \gewnu with high accuracy and processes
of the third and fourth order (multiple gauge boson production),
in particular $e\gamma\to eWW$.  In the processes $\egam\to
eWWZ$, $\ggww WW$, $\ggww ZZ$, one can study subprocesses with
heavy gauge boson scattering.

2) Hunting for effects from strong interaction in the Higgs
sector.

\vspace{0.3cm}

{\em This work is supported by grants of International Science
Foundation ISF RPL300, of INTAS -- 93 -- 1180 and of Russian Fund
of Fundamental Investigations RFFI -- 93--02--03832}.

{\large\bf Appendix. Anomalies and RC}\\

The RC with anomalies are divergent. To kill these divergencies,
one should introduce counter-terms of higher orders. Therefore,
the Lagrangian with operators of dimension, e.g., up to 6 should
be added by items of 8-h order, etc. (That is standard picture
for the nonrenormalizible theory). It means that the calculation
of RC with anomalies is beyond the accuracy of basic approximarion
with effective Lagrangian.

In practice, one can introduce some formfactors of anomalies. The
values of RC depend on the scale of these formfactors. Therefore,
the effective Lagrangian should be added by some additional
prescriptions that include both the scales of anomalies and the
relation between these scales in different "directions".
Besides, some additional constant items from operators of higher
degrees and dimensions can vary results strongly.

Therefore, {\em (i)} the RC are out of accuracy of approximation
used when anomalous interactions taken into account. The
calculations with anomalies have sense if only these anomalies
are large enough to standard RC could be invisible in comparison
with anomalies.  {\em (ii)} The calculation of RC with anomalies
could have sense if only we know much enough about the structure
of underlying theory, for which the EW theory is low energy
limit.

\section*{Figure captions}

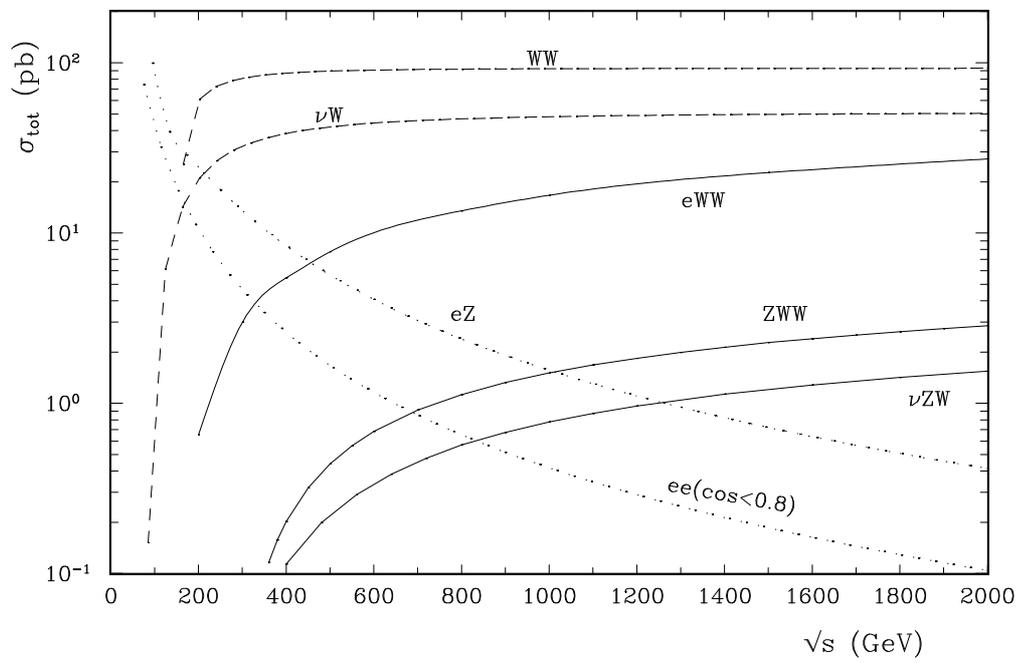
\begin{figure}[htb]
\caption{ The cross sections of some processes in \ggam and \egam collisions.}
\label{Fig1}
\end{figure}

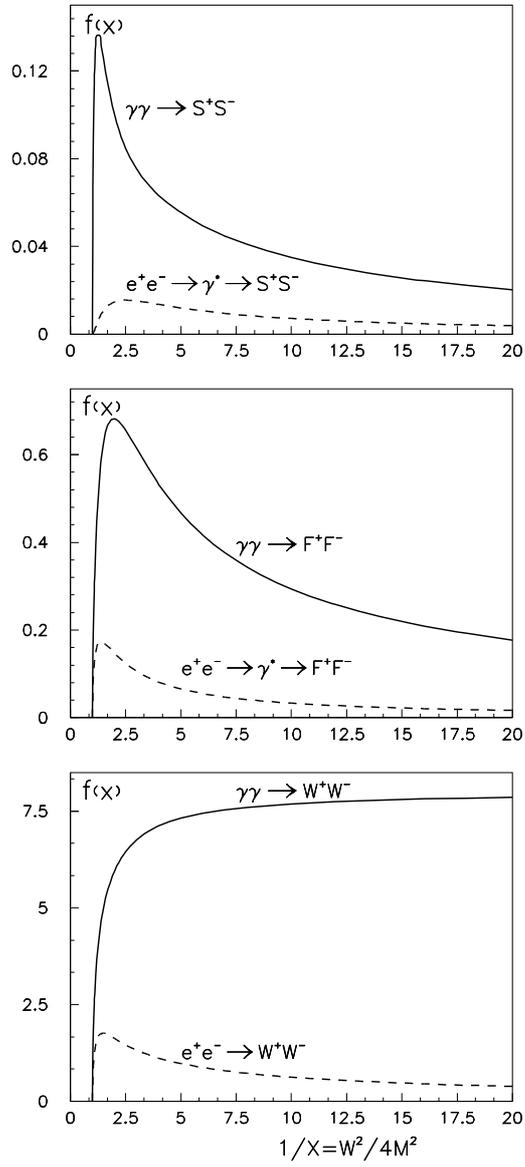
\begin{figure}[htb]
\caption{ The cross sections for charged pair particles (P)
production in \ggam and \epe collisions, $\sigma
=(\pi\alpha^2/M^2)f_P(x)$, where M is particle mass, $x=s/4M^2$.
The functions $f_P(x)$ are shown for P=S (scalars), F (fermions)
or W (vector gauge bosons).}
\label{Fig2}
\end{figure}

\end{document}